\PassOptionsToPackage{unicode=true}{hyperref} 
\PassOptionsToPackage{hyphens}{url}
\documentclass[]{article}
\usepackage{lmodern}
\usepackage[superscript,biblabel]{cite}
 \usepackage{tikz} 
\usepackage{amssymb,amsmath}
\usepackage{ifxetex,ifluatex}
\ifnum 0\ifxetex 1\fi\ifluatex 1\fi=0 
  \usepackage[T1]{fontenc}
  \usepackage[utf8]{inputenc}
  \usepackage{textcomp} 
\else 
  \usepackage{unicode-math}

  \defaultfontfeatures{Ligatures=TeX,Scale=MatchLowercase}
\fi
\IfFileExists{upquote.sty}{\usepackage{upquote}}{}
\IfFileExists{microtype.sty}{%
\usepackage[]{microtype}
\UseMicrotypeSet[protrusion]{basicmath} 
}{}
\IfFileExists{parskip.sty}{%
\usepackage{parskip}
}{
\setlength{\parindent}{0pt}
\setlength{\parskip}{6pt plus 2pt minus 1pt}
}
\usepackage{hyperref}
\hypersetup{
            pdftitle={Blended Survival Curves: A New Approach to Extrapolation for Time-to-Event Clinical Trial DATA in Health Technology Assessment},
            pdfborder={0 0 0},
            breaklinks=true}
\usepackage[margin=1in]{geometry}
\usepackage{longtable,booktabs}
\IfFileExists{footnote.sty}{\usepackage{footnote}\makesavenoteenv{longtable}}{}
\usepackage{graphicx,grffile}
\makeatletter
\def\maxwidth{\ifdim\Gin@nat@width>\linewidth\linewidth\else\Gin@nat@width\fi}
\def\maxheight{\ifdim\Gin@nat@height>\textheight\textheight\else\Gin@nat@height\fi}
\makeatother
\setkeys{Gin}{width=\maxwidth,height=\maxheight,keepaspectratio}
\ifx\paragraph\undefined\else
\let\oldparagraph\paragraph
\renewcommand{\paragraph}[1]{\oldparagraph{#1}\mbox{}}
\fi
\ifx\subparagraph\undefined\else
\let\oldsubparagraph\subparagraph
\renewcommand{\subparagraph}[1]{\oldsubparagraph{#1}\mbox{}}
\fi

\makeatletter
\def\fps@figure{htbp}
\makeatother

\usepackage{bm}
\usepackage{caption}
\usepackage{subcaption}

\title{Blended Survival Curves: A New Approach to Extrapolation for Time-to-Event Outcomes from Clinical Trial in Health Technology Assessment}
\author{Zhaojing Che\footnote{Research was funded by UCL-CSC joint doctoral training grant.}, Nathan Green and Gianluca Baio \\\\\\
Department of Statistical Science, University College London\\
}

\date{}

\begin{document}

\maketitle

\begin{abstract}

\noindent \textbf{Background} Survival extrapolation is essential in the cost-effectiveness analysis to quantify the lifetime survival benefit associated with a new intervention, due to the restricted duration of randomized controlled trials (RCTs). Current approaches of extrapolation often assume that the treatment effect observed in the trial can continue indefinitely, which is unrealistic and may have a huge impact on decisions for resource allocation. \textbf{Objective} We introduce a novel methodology as a possible solution to alleviate the problem of performing survival extrapolation with heavily censored data from clinical trials. \textbf{Method} The main idea is to mix a flexible model (e.g., Cox semi-parametric) to fit as well as possible the observed data and a parametric model encoding assumptions on the expected behaviour of underlying long-term survival. The two are “blended” into a single survival curve that is identical with the Cox model over the range of observed times and gradually approaching the parametric model over the extrapolation period based on a weight function. The weight function regulates the way two survival curves are blended, determining how the internal and external sources contribute to the estimated survival over time. \textbf{Results} A 4-year follow-up RCT of rituximab in combination with fludarabine and cyclophosphamide v. fludarabine and cyclophosphamide alone for the first-line treatment of chronic lymphocytic leukemia is used to illustrate the method. \textbf{Conclusion} Long-term extrapolation from immature trial data may lead to significantly different estimates with various modelling assumptions. The blending approach provides sufficient flexibility, allowing a wide range of plausible scenarios to be considered as well as the inclusion of genuine external information, based e.g. on hard data or expert opinion. Both internal and external validity can be carefully examined.
    
\end{abstract}

\hypertarget{introduction}{%
\section{Introduction}\label{introduction}}

In health economic evaluations, survival outcomes (or other time-to-event data) from randomised control trials (RCTs) are typically used to assess cost-effectiveness of new interventions.  However, the observed data from RCTs are often censored and immature with limited duration of follow-up \cite{tai2021prevalence}, so the clinical benefits regarding  life expectancy or quality-adjusted life-years (QALYs) cannot be obtained directly. Consequently, it is necessary to ``extrapolate'' the estimates of the resulting survival proportions, often long beyond the data observed in the trial period \cite{bell2019review}.

Methods of extrapolation most often used in submissions to health technology agencies such as the National Institute for Health and Care Excellence (NICE) in the UK, often consider a parametric model for the control arm and assume proportional hazard (PH) to derive the survival curve for the treatment arm \cite{ball2021onwards, kearns2020uncertain, benedict2018survival}. This implicitly assumes a constant treatment effect beyond the trial period. However, a treatment performing well over the course of the trial is unlikely to keep consistent on account of various factors such as waning effect of treatment or competing risks from other cause of mortality. The typical length of follow-up in clinical trials has been shown to account for no more than $40\%$ of the modelled time horizon \cite{gallacher2019pharmaceutical}, not reaching the median time. In the absence of long-term data, care should be taken in whether the extrapolation is realistic, as the long-term modelling assumptions can have a dramatic impact on the decisions  \cite{everest2021parametric, everest2022parametric}.

Historically, conventional approaches involved fitting the most appropriate parametric model to the observed data \cite{jackson2010survival}. In fact, different models with a similar fit to the data may generate highly divergent long-term survival estimates owing to the differences in the tails of survival distributions. Recently, there has been an increasing recognition that external long-term validity is essential when the extrapolation period is substantial with heavy censoring in the trial data \cite{kearns2020uncertain, kearns2021comparing, bullement2022incorporating}. Current guidelines recommend the inclusion of both statistical criteria for model fitting as well as clinical plausibility of extrapolation, which may be achieved through the use of external data or expert opinion \cite{rutherford2020nice}. In recent time then, the proportion of health technology appraisals (HTAs) using external information for validity has increased sharply \cite{gallacher2019pharmaceutical, bell2019review}, in which clinical experts assess the plausibility of extrapolation or evaluate which models fall in with the elicited plausible range of survival \cite{gibson2017modelling}.  

 There are many different ways that external data can be leveraged \cite{kearns2021comparing, Jackson2017}. While the most frequent methods are indirect or retrospective, direct utilisation of patient-level data for the extrapolation have increasingly been considered  \cite{gibson2017modelling, rutherford2020nice}. It is possible that historical data are formally integrated into the extrapolated portion as informative priors via a Bayesian framework \cite{soikkeli2019extrapolating}. In addition, a piecewise or hybrid approach where observational data are used to facilitate the extrapolation has been undertaken, though selection of where to implement cut points can be fairly subjective or arbitrary \cite{bullement2022incorporating, national2012abiraterone, larkin2015predicted}. Since longer term registry data are commonly only available for the control arm this may not provide specific knowledge regarding the treatment arm. A constant or decreasing HR was frequently suggested for the intervention group \cite{national2012Rituximab, national2007bortezomib, national2008pemetrexed}. Including general population mortality into a mixture cure model also provides a chance to improve the accuracy of extrapolations \cite{kearns2021extrapolation,  felizzi2021mixture}. Further methods were attempted to combine available evidence on general population survival, cancer registry data, or clinical opinion \cite{guyot2017extrapolation, benaglia2015survival, cope2019integrating}.   

This paper presents a method based on ``blending'' survival curves as a possible solution. A similar blending approach has been presented previously in other applied fields \cite{castro2021practical} but we modify this idea to survival modelling for cost-effectiveness analysis. The basic idea is to mix a flexible model (e.g. Cox semi-parametric) to fit as well as possible the observed data and a parametric model encoding assumptions on the expected behaviour of underlying long-term survival. The blended curves will improve decision making especially in the cases where
decisions are made accounting for survival in long-term timeframes relative to the available trial data
but expert knowledge or external data about the long-term is available and can be coherently combined.
Extrapolated curves using the short-term data only are likely to be bias/over-estimate survival
so with the blended model we can constrain the tail and retain the information in the early time period.
For HTA, cost and QALYs calculations can use the estimated survival in the blending interval which is consistent with information from both the early and later stages.

\hypertarget{introduction}{%
\subsection{Motivating Case Study}\label{Motivation}}

Our motivating example is the one considered in NICE technology appraisal TA174 \cite{national2009rituximab} and in other methodological contributions \cite{williams2017estimation, williams2017cost}. This is based on the CLL-8 trial \cite{hallek2010addition}, which compares rituximab with fludarabine and cyclophosphamide (RFC) to fludarabine and cyclophosphamide (FC) for the first-line treatment of chronic lymphocytic leukemia. 

Among 810 patients enrolled in the trial, 403 were randomly assigned to receive the treatment of RFC and the remaining 407 to the control arm of FC. There were 41 and 52 deaths in the RFC and FC arms respectively \cite{williams2017estimation}. While this study has a relatively large sample size and a relatively long follow-up (around 4 years), it is also characterised by a large amount of censoring such that over 70$\%$ of individuals were not observed to die, as is common in this type of investigations.

\begin{figure}[!h]
    \centering
    \input{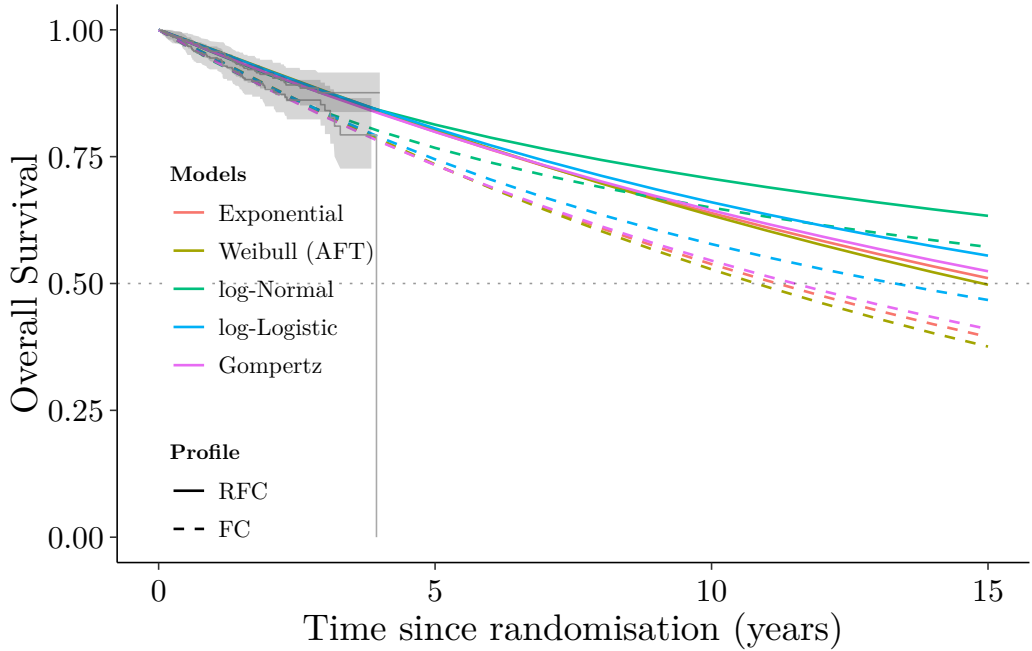}
    \caption{Overall Survival curves for the parametric models (Exponential, Weibull, log-Normal, log-Logistic, Gompertz) fitted to the 4-year CLL-8 trial data (Kaplan-Meier curves) and long-term extrapolation to  15 years.}\label{fig:data}
\end{figure}

Following existing guidance \cite{rutherford2020nice, latimer2011nice}, a set of standard parametric distributions were fitted to the digitised data on overall survival from published Kaplan-Meier curves, as shown in Figure~\ref{fig:data}. These models achieved a feasible fit to the observed data (as evident in the left portion of Figure~\ref{fig:data}), but none of them generated credible extrapolations. All models suggested over 30\% survival at 15 years, which was in stark contrast with expert estimates, suggesting instead that only 1.3\% of the cohort would be likely to survive beyond that time \cite{roche2008rituximab}.

We use this case study to illustrate how the blended curve methodology may help alleviate similar issues with extrapolation of heavily censored survival curves.

\hypertarget{methodology}{%
\section{The blended curve method}\label{methodology}}

Denote the observed data as $\mathcal{D}_i=(t_i, d_i)$, where $t_i$ is the observed time at which the event (e.g.~progression or death) occurs, while $d_i$ is an event indicator taking value 1 if the $i-$th individual is fully observed and 0 if censored. Typically, we model $t_i\mid\bm{\theta}_i \sim p(t\mid\bm\theta)$, where $p(\cdot)$ is a parametric distribution indexed by a vector of parameters $\bm\theta$, for instance $\bm\theta=(\gamma,\mu)$ indicating shape and scale, respectively. Given this structure, we can define the hazard $h(t\mid\bm\theta)$ and the survival function $S(t\mid\bm\theta)=\Pr(T>t\mid\bm\theta)$. 

The ``blending'' idea is to consider two separate processes to describe the long-term horizon survival. The first one is driven exclusively by the observed data. Similar to a ``standard'' HTA analysis, we use this to determine an estimate over the entire time horizon, which we term $S_{obs}(t\mid\bm\theta_{obs})$, a function of the relevant parameters $\bm\theta_{obs}$.
We could choose a simple parametric model or, alternatively, some other more complex model, with the main objective to produce the \textit{best} fit possible to the observed information. Unlike in a ``standard'' modelling exercise where the issue of overfitting is potentially critical, achieving a very close approximation to the observed dynamics has much less important implications in the case of blending, as explained further below. 

For the second component of the blending process, we consider a separate ``external'' survival curve, ${S_{ext}(t\mid\bm\theta_{ext})}$. This is a parametric model that is not informed by the observed data --- for instance, we could use ``hard'' information, e.g.~derived from a different data source (such as registries or observational studies), or construct a model that is purely based on subjective knowledge elicited from experts, or possibly a combination of the two. Either way, $S_{ext}(t\mid\bm\theta_{ext})$ will typically be less concerned with the observed portion (for which we want the available data to drive the inference), but is instrumental to produce a reasonable and realistic \textit{long-term} estimate for the survival probabilities.

The ``blended'' survival curve is simply obtained as 
\begin{align}
    S_{ble}(t\mid\bm\theta) = S_{obs}(t\mid\bm\theta_{obs})^{1-\pi(t; \alpha, \beta, a, b)}\times S_{ext}(t\mid\bm\theta_{ext})^{\pi(t;\alpha, \beta, a, b)}
    \label{ble_for}
\end{align}

where $\bm \theta = \{\bm \theta_{obs}, \bm \theta_{ext}, \alpha, \beta, a, b\}$ is the vector of model parameters. Here, $\pi(\cdot)$ is a weight function that controls the extent to which the two survival curves $S_{obs}(\cdot)$ and $S_{ext}(\cdot)$ are blended together. Technically, we define $\pi(\cdot)$ as the cumulative distribution function of a Beta random variable with parameters $\alpha,\beta > 0$, evaluated at the point $(t-a)/(b-a)$
\begin{align*}
    \pi(t;\alpha,\beta,a,b) = \Pr\left(T\leq \frac{t-a}{b-a}\mid \alpha,  \beta\right) = F_{\text{Beta}}\left (\frac{t-a}{b-a}\mid \alpha, \beta \right),
\end{align*}
for $t \in [0,T^*]$, where $T^*$ is the upper end of the interval of times over which we want to perform our evaluation.
This means that the weighting function $\pi(\cdot)$ varies over the time horizon, which in turn allows us to give different weights to the two components at different times $t$.  The range $[a, b]\in (0,T^*)$ is the \textit{blending interval}, i.e.~a subset of the life-time horizon in which $S_{obs}(\cdot)$ and $S_{ext}(\cdot)$ are blended into a single survival curve. 

\begin{figure}[!h]
     \input{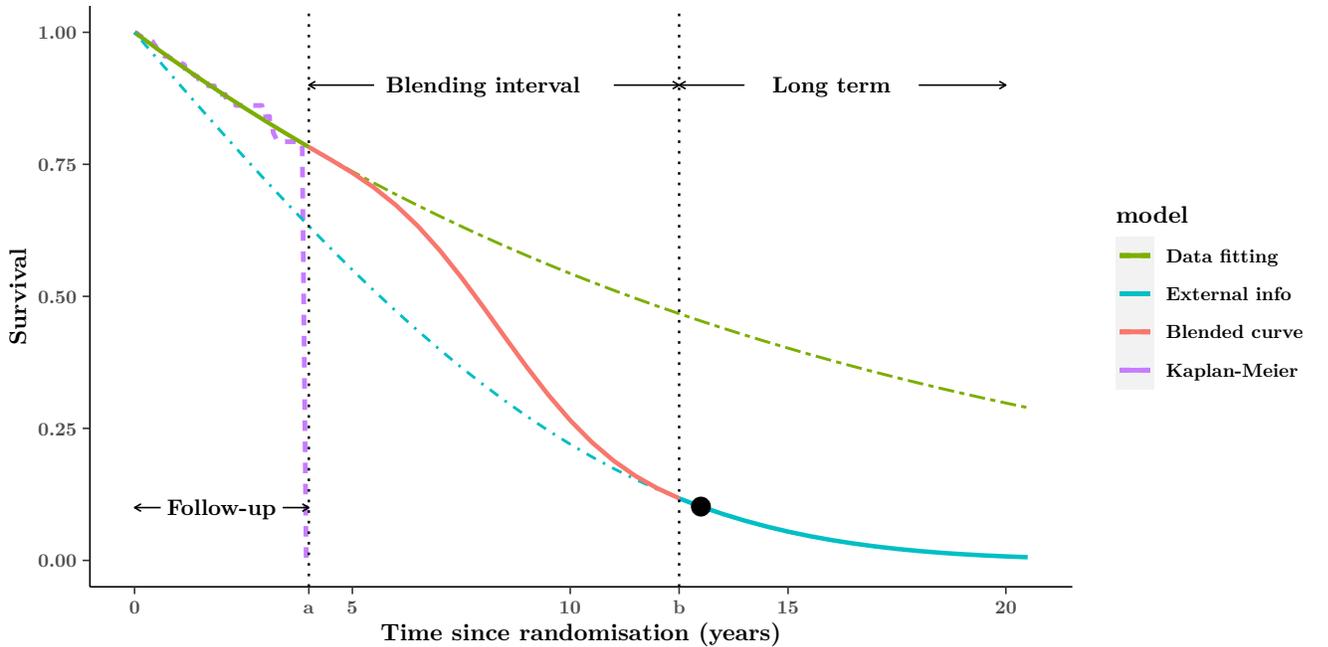}
     \caption{Graphical representation of the blended curve method. The whole time-horizon is partitioned into three parts: Follow-up, Blending interval and Long term. The blended survival is equivalent to the model fitted to the short-term data (purple KM curve) within Follow-up period (green curve); then gradually approaching the external estimate in the Blending interval (red curve); eventually consistent with the expected behaviour (blue curve) in the Long term. The black point in the Long term is an example of external information about 10\% expected survival at the 13 years from experts.}\label{blending}
 \end{figure}

Figure~\ref{blending} depicts this process graphically. In this case, we assume that the trial data span over the interval $[0,a]$, which we label in the graph as the ``Follow-up''. The dashed curve is the Kaplan-Meier (KM) estimate of the observed data (for simplicity,  but without loss of generality, we consider here a single arm). The curve labelled as $S_{obs}$ results from a suitable model fitted to the observed data, in order to capture the known features of the data generating process almost to perfection --- as is possible to appreciate in the graph, the KM curve is basically identical with the model obtained with $S_{obs}$.

The blue curve, indicated as $S_{ext}$ should be used to give information about the expected long-term behaviour of the survival process. While it may be difficult to directly access hard data, as discussed in Section \ref{ext}, we and others \cite{Jackson2017, Uncertain} argue that it is often possible and generally desirable to so. For example, we may have individual level data from a registry based on a drug with a similar mechanism to the one of interest; or perhaps we have elicited clinical knowledge or expert opinion to identify that survival at a certain time point is not expected to exceed a certain threshold and we can use this information to constrain $S_{ext}$ to conform with this expectation. Notice in particular that $S_{ext}$ can deviate substantially from the observed data, as shown in Figure~\ref{blending}. 

As mentioned above, the interval $[a,b]$ is the portion of the lifetime horizon in which the blending process occurs. In other words, $S_{ble}(\cdot)$ is constructed as a combination of $S_{obs}(\cdot)$ and $S_{ext}(\cdot)$ where:

\begin{itemize}
    \item Between times 0 and $a$, $\pi(\cdot)=0$, which means that the long-term extrapolation has no influence. Since this is the trial follow up, the observed data should be described as best as possible, as obtained by $S_{obs}(\cdot)$.
    \item Between times $b$ and $T^*$ (set to 20 in the example shown in Figure~\ref{blending}), $\pi(\cdot)=1$, which means that it is the \textit{long-term} extrapolated survival curve from the observed data to bear no weight whatsoever. Again, we do this because, given the heavy censoring, the resulting extrapolation is most likely a gross overestimation.
    \item Between times $a$ and $b$, the two curves merge into one another, according to the process characterised by the weight function $\pi(\cdot)$. In the ``blending interval'', the two curves both influence the resulting blended survival curve, which gradually abandons the extrapolation from the observed data (thus avoiding issues with the inherent overfitting and unrealistic estimates) and merges into the long-term extrapolation from the external evidence. 
\end{itemize}

\begin{figure}[!h]
     \centering
     \input{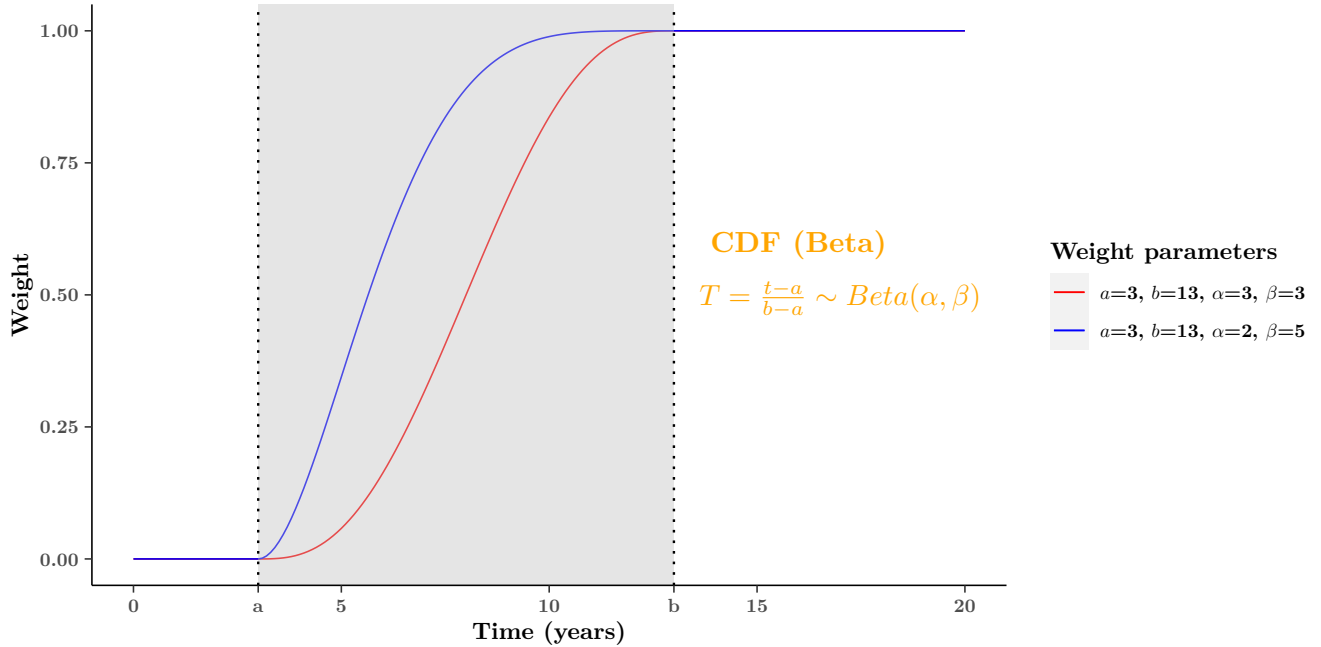}
     \caption{Graphical examples of the weight function \(\pi(t; \alpha, \beta, a, b)\). The grey area $[a, b]$ is the blending intervals $[3, 13]$ for both two weight curves. Slope of the red curve ($\alpha=3$, $\beta=3$) is greater than the blue one ($\alpha$=2, $\beta =5$) , which means the former blending rate is slower than the latter one.   }\label{beta}
 \end{figure}

 We can control the rate at which the blending process occurs by using specific values for the parameters ($\alpha,\beta)$ of the relevant Beta distribution. 
Given the same blending area, different values of  parameters in the Beta distribution will provide distinct slopes, influencing the speed of the blending process. For example, in Figure~\ref{beta}, in the same blending interval $(a=3, b=13)$, the blue curve $(\alpha=2, \beta=5)$ is steeper than the red one $(\alpha=\beta=3)$, which implies that the blending trend of the former is faster and the impact of $S_{ext}$ would be relatively greater at the same point in time (along the $x-$axis). Overall, the slope of the weight curve in the situation that $\alpha < \beta$ is larger than the alternative circumstance that $\alpha \geq \beta$. 

Different assumptions about how quickly the treatment effect might wane off can be easily examined by adjusting the choice of parameters regarding the weight function  $\pi(t)$ as a part of sensitivity analysis. For example, if the observed
treatment effect is assumed to persist over the whole horizon, we can set the value of $a$ equal to the point $T^{*}$, in which case the blended curve is the same as the observed one over the entire timeframe.

Note also that our method is fundamentally different from well established mixture cure models (MCMs;  \cite{lambert2007estimating}). In the MCM case, it is assumed that the observed trial data correspond to a mixed survival curve resulting from the experience of two subgroups (``cure’’ vs ``non-cured’’ patients). Conversely, we model two components $S_{obs}$ and $S_{ext}$ independently within the blended process, respectively based on observed data and external evidence. Importantly, values for $\pi(t)$ are provided externally and could be modified on demand. We return to this important distinction in Discussion (Section~\ref{dis}).

\hypertarget{methodology}{%
\subsection{Blending hazard functions}\label{haz}}

By simply rescaling Equation (\ref{ble_for}), our method can also be expressed in terms of hazard functions. This is helpful because hazard plots often aid understanding of long-term survival mechanism and provide useful insights into suitable model selection  \cite{bell2019review}. Specifically, the blended hazard rate $h_{ble}(t)$ can be characterised by three components: the weighted hazard rates from two survival curves $h_{obs}(t)$ and $h_{ext}(t)$ and an extra term related to the weight function and cumulative hazard. Then, we can re-express Equation (\ref{ble_for}) equivalently as
 \begin{align*}
    h_{ble}(t) = \Big(1-\pi (t)\Big) \times h_{obs} (t) + \pi (t)\times h_{ext} (t) + \frac{f_{\text{Beta}}(\frac{t-a}{b-a})}{b-a}\times \Big( H_{ext}(t) - H_{obs}(t) \Big), 
\end{align*}
where the term $f_\text{Beta}(\cdot)$ denotes the density function of a Beta random variable, associated with the weight function $\pi(\cdot)$, while $H_{ext}(t)$ and $H_{obs}(t)$ are the cumulative hazard rates from the two underlying survival curves, respectively. 

The hazard function depends on the same subset of parameters as the corresponding survival functions. Given the properties of the Beta distribution, $f_{Beta}(\cdot)$ only supports the blending interval (i.e., $[a, b]$), but is zero, otherwise. Since $\pi(t)$ is 0 in $[0, a)$ and 1 in $[b, T^{*}]$, it is easy to show that the blended hazard $h_{ble}$ is equal to the observed estimate $h_{obs}(t)$ at the beginning, and to the external hazard $h_{ext}(t)$ in the long term after time point $b$.

\begin{figure}[h]
    \centering
    \input{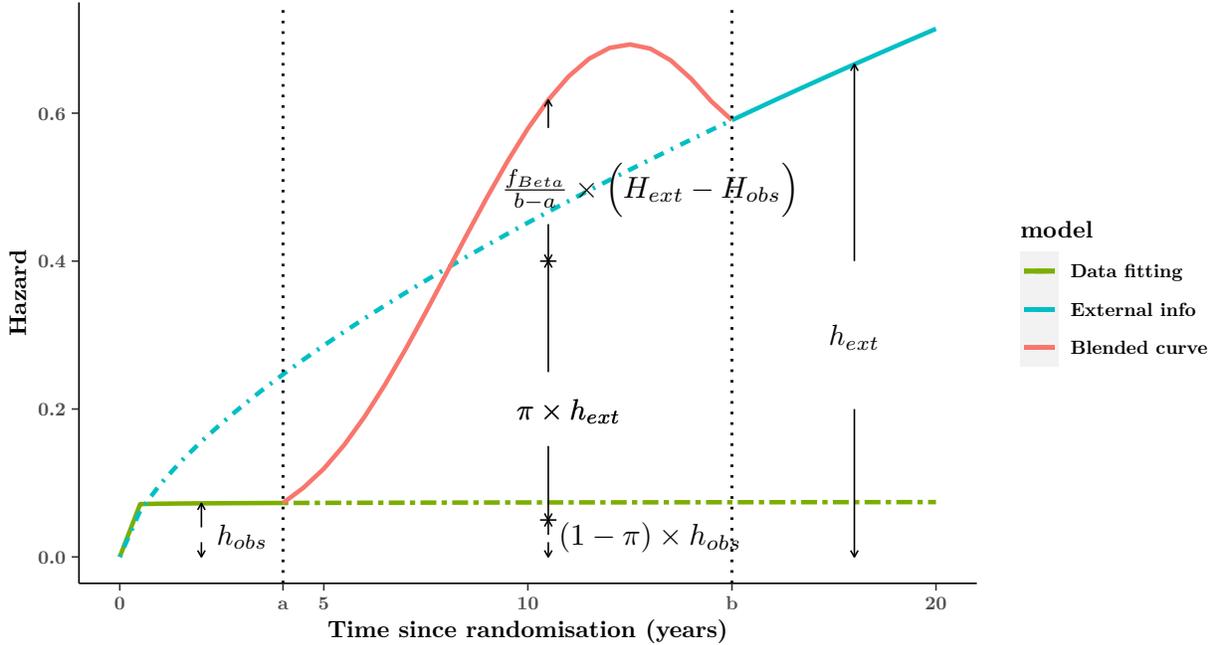}
    \caption{Graphical representation of the blended hazard. For interval $[0, a]$, the blended hazard is equal to the observed hazard ($h_{obs}$, green curve); then in the blending interval $[a, b]$, there is a sharp increase followed by a steady decrease ($h_{ble}$, red curve); eventually, it is consistent with the external hazard ($h_{ext}$, blue curve).}\label{haza}
\end{figure}

The slope of the blended hazards as well as the location of the turning point can be determined by the value imposed for the parameters $\alpha,\beta$, enabling different assumptions on the underlying hazard rates to be tested. In this example, in which the external hazard is much greater than the observed hazard, if $\alpha<\beta$, there would be monotonic increasing hazard within the interval $[a, b]$. Alternatively, if $\alpha>\beta$, a sharp increase would be followed by a steady decrease during the interval (red segment in Figure \ref{haza}). This pattern allows the turning points beyond the observed period and so it is likely that form of the blended hazard is more flexible and realistic compared to standard parametric models. 

\hypertarget{Technical work}{%
\section{Technical implementation of the blended model}}

\hypertarget{Technical work}{%
\subsection{Observed time period: \textit{best} fit to the internal data}\label{obs}}

Generally speaking, there is no restriction to the distributional assumptions used to model the observed data. With a view to providing the best fit possible and a good level of flexibility, here we recommend a Cox semi-parametric model with piecewise constant hazards. We choose a Bayesian approach, which naturally allows the incorporation of external evidence and lends itself to the conduct of ``uncertainty analysis'' \cite{baio2013bayesian, baio2015probabilistic}.

To construct the model, we partition time period into $K$ intervals, $0=u_{0}<u_{1}<u_{2}\cdots<u_{K}$, and assume the baseline hazard $h_{0}(t)$ to be constant in each interval using $K$ parameters $\lambda_{1}$, $\dots$, $\lambda_{K}$. We model the parameter $\lambda_{k}$ with random walks (RW) of order one (or two) as the prior assuming that increments $\Delta \lambda_{k} = \lambda_{k}-\lambda_{k-1}$ (or $\Delta^{2} \lambda_{k} = \lambda_{i}-2\lambda_{i-1}+\lambda_{i-2}$) follow a Gaussian distribution with zero mean and a common precision \cite{gomez2020bayesian}. 

Note that using this model, we can still extrapolate beyond the observed times using the RW structure. Obviously, in the presence of large censoring, the extrapolation is likely to be incredible with substantial uncertainty around the average. This, however, is a minor concern in our modelling structure, because as time progresses outside of the blending interval, the extrapolation from the semi-parametric component has increasingly low weight.

Of course, other choices are possible: we could select a parametric model (e.g.~Weibull, Gompertz, or any other from the set suggested in various guidelines \cite{bell2019review}) --- in reality, a flexible semi-parametric model may not increase the computational complexity by a substantial amount, compared to alternatives such as Royston-Parmar splines \cite{royston2002flexible} or fractional polynomials \cite{royston1994regression}. In addition, because of the blending process, we only need to worry about the performance of any model chosen in the ``follow up''~period.

\hypertarget{Technical work}{%
\subsection{Unobserved time period: extrapolation using external data}\label{ext}}

In the best case scenario, long-term data can be accessed from a relevant study, possibly of an observational nature, such as a registry or a cohort study; this is naturally unlikely to contain direct information on the intervention under investigation from the trial data. But perhaps, we may have information on drugs with similar mechanisms of action, or tackling the same condition. In these circumstances, we could simply perform a ``standard’’ analysis of the relatively complete data, using a parametric model. Whatever the distributional assumptions, we would be able to determine an estimate of either the survival or the hazard curve for the extrapolation (long-term) period and then plug that into the blended model.

\hypertarget{Technical work}{%
\subsection{Unobserved time period: extrapolation using expert judgement}\label{ex}}

A more general situation, encountered in real-life applications, is when only tentative knowledge is available, typically in the form of expert elicitation. It is rather common for modellers to ask ``key opinion leaders'' for their assessment of the validity of a given extrapolation, perhaps in the form of plausible ranges or point estimates for the survival probabilities at given times. For example, experts may suggest that, given their clinical knowledge, the plausible interval of 10 year survival probability is between 10$\%$ and 30$\%$, or that no more than 5$\%$ of participants would survive beyond 15 years. We thus need to map those numerical estimates onto a suitable model and construct a representative curve of the external information. 

Elicitation of survival estimates could be expressed as the expected number of individuals who, in a population of a given size, will survive at the specific point; for instance, 20$\%$ survival at 10 years could be interpreted as ``20 in 100 patients would survive beyond 10 years''. We could translate the clinical constraint into an artificial dataset and then use standard method to analyze the pseudo data. Given that 80$\%$ of time-to-event data should be smaller than 10 years, we could use a Uniform distribution with boundaries 0 and 10 to generate the individual survival times, because there is no knowledge or assumption about the time-to-event outcome in this synthetic dataset. To build up survival outcomes for the remaining 20\% of the population who would survive at least 10 years, it is essential to determine a maximum lifetime $T_{max}$ beyond which no patient would be expected to be alive, and then similarly, the survival times should be the samples of the Uniform distribution ranging from 10 years to $T_{max}$. Of course, other processes of simulation of the underlying time to event data may be selected, as long as the soft-constraints hold and the resulting long-term extrapolation is justifiable.

\begin{figure}[h]
    \centering
    \includegraphics[width = 10cm]{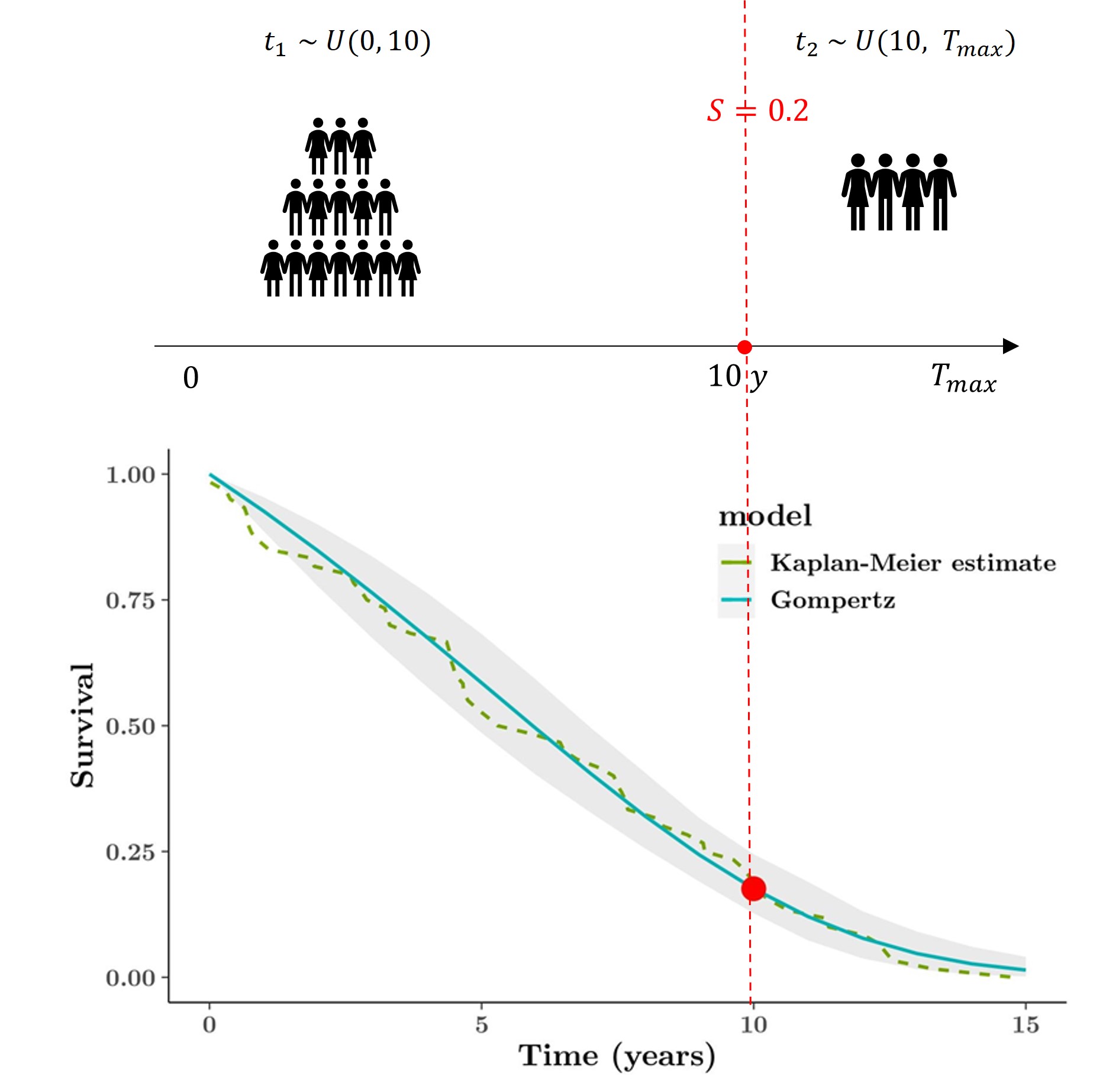}
    \caption{Graphical representation of constructing external survival curve (based on the subjective opinion). The top plot illustrates the mechanism of generating the artificial dataset and the bottom plot is the Gompertz model fitted to the synthetic dataset. The elicitation is only at one time point: 20\% expected survival at 10 years.}\label{eb}
\end{figure}

Figure \ref{eb} (top plot) illustrates the above example and the synthetic dataset consists of the two groups of time-to-event data $t_1$ and $t_2$, in which all the event indicators are equal to 1 as they are assumed to be fully observed.  The dashed curve in Figure \ref{eb} (bottom plot) shows the Kaplan-Meier estimate for the synthetic dataset with 100 individuals. In fact, the choice of the sample size is directly related with the level of implied uncertainty on the external information. If clinicians/experts are not very certain about their elicitation, the sample size of the artificial dataset should be reduced, which would lead to a wider 95$\%$ interval around the point estimate. When the dataset is constructed, the process is similar to the one used for direct ``hard'' external data. In Figure \ref{eb}, fitting a Gompertz distribution appears to perform well and the blue curve is fully reflective of the expert opinion at 10 years. 

This simple case only considers one time constraint, but the mechanism for multiple time points would be essentially identical and effected through a partition of the time horizon into three or more portions. In that case, the resulting curve can align more closely with substantive expert beliefs.

\hypertarget{results}{%
\section{Results}\label{res}}

\hypertarget{results}{%
\subsection{Interim analysis for observed time period}}

\begin{figure}
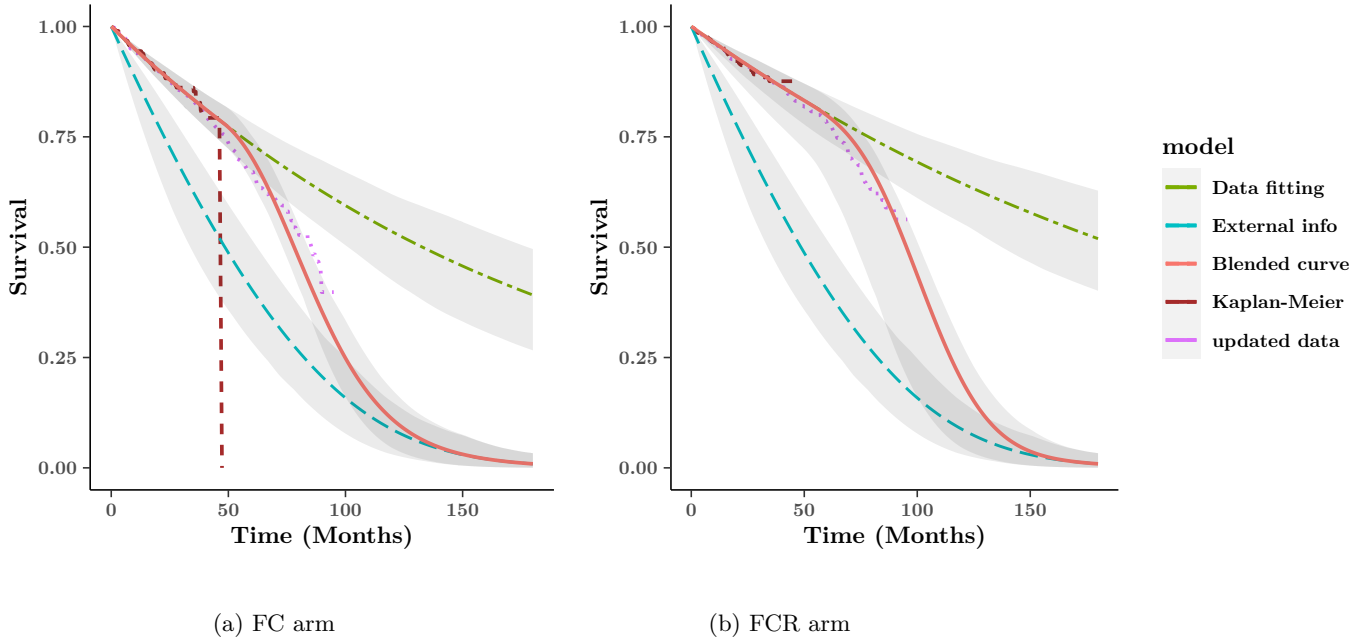

              \begin{subfigure}{.4\textwidth}
                    \hspace*{-0.5cm}
		\input{Figure7.tex}
		\caption{FC arm}
                   \end{subfigure}
 \begin{subfigure}{.4\textwidth}
                    \hspace*{0.5cm}
		\input{Figure6.tex}
		\caption{FCR arm}
                   \end{subfigure}

	 \caption{Blended survival curve based on short-term data and external information for RFC arm. The digitised data from CLL-8 are updated with longer follow-up till 96 months (purple dotted line).}
	 \label{up}

	\end{figure}

 The piece-wise constant hazard model in Bayesian framework provided a good fit to the observed data (with 8 intervals over the 4-year follow-up; green curve in Figure~\ref{up}). As is known, a greater number of intervals might lead to lower deviance (\textit{better} fit); however, in this particular case, no meaningful improvement was seen by increasing the number of intervals. Notice that, unsurprisingly, the extrapolation from the model is not reasonable, as it implies artificially and unrealistically large survival probabilities at the end of the follow-up period.

\hypertarget{results}{%
\subsection{External curve with expert information}}

Given the relatively strong opinion that approximately 1.3\% of the cohort would be alive beyond 15 years, we construct a synthetic dataset with 300 participants, in which no more than four individual times are longer than 15 years (180 months). We can experiment with different sample sizes (in our case, we used a number of scenarios, with sample size ranging from 10 to 500) to get a better sense of the implied uncertainty around the resulting survival curves.

Among the candidate parametric models, the Gompertz distribution fits the external data very well, describing the belief specified above accurately. Since the external information is assumed to be rather certain, we imply a relatively narrow 95\% interval estimate in Figure \ref{up}, ranging between 0.5\% and 2\%.  In a real-life case, the experts and modellers should be able to defend this assumption, in the absence of hard evidence to justify it. We note, however, that this process happens irrespective of the modelling strategy chosen --- in our case, we make it in a way that does not affect directly the fit to the observed data.

\hypertarget{results}{%
\subsection{Blended estimate compared to updated data from CLL-8 trial}}

Figure \ref{up} shows the blended survival estimate driven by internal and external curve over the whole time-horizon. Without any further information about the blending process, we assume a constant speed over the blending interval, based on a linear weight function with $\alpha=\beta=1$. On account of the only elicited time point at 15 years, we identify the blending interval from the end of follow-up (4 years) to the end of the modelling horizon.

When compared to a later data-cut for the CLL-8 trial till 96 months \cite{fischer2016long}, the blended survival curve after 48 months is generally very close to the updated data. Unsurprisingly, the observed survival without external information overestimate the longer value, 40\% higher than the updated result.

\hypertarget{discussion}{%
\section{Discussion}\label{dis}}

There is a growing need to improve extrapolation of immature survival data when interim analysis is frequently carried out in the context of accelerated regulatory approvals. A short duration of follow-up is often subject to a substantial amount of censoring, which can lead to implausible extrapolations with conventional approaches only based on observed data. In addition, innovative cancer drugs are evaluated on the back of limited information, because no alternative treatment is available as a viable option for patients affected by a specific disease. To obtain credible estimation of overall survival gains, it is essential to relax the traditional PH assumption and supplement the external information to guide the extrapolated curve. In this paper, we have introduced an innovative approach based on blended survival curves as a possible solution to these issues in the extrapolation.

In the cases when the hazard early on is unlikely to reflect the long-term behaviour, our blended approach enables the extrapolated survival to be less and less affected by the short-term data as time progresses. Long-term outcomes would be dominated by the external information. Providing a best fit to the observed data, the blended curve would gradually approach the prediction derived from the external sources over the extrapolated period. In the blending interval, time-specific weights are allocated to the observed and external survival that allow for different proportions of the two contributing components to the overall estimate. As mentioned in Section~\ref{methodology}, a mixture cure model also consists of two components (survival profiles of cured and uncured patients), but is distinct from the blended model because it assumes a constant weight, namely the proportion of cured patients, through the entire time range \cite{andersson2011estimating}. In addition, cure fractions- as well as the survival of uncured patients- often solely rely on short term data \cite{latimer2022extrapolation}, while weight functions, together with external projections in the blended curves, would be governed by information outside an RCT. Finally, MCMs are based on the assumption that the underlying data generating process gives rise to a single survival mechanism that is a combination of two subgroups; in our case, we explicitly consider two separate process (the short-term and long-term survivals) and ensure that extrapolation from the former is anchored in a principled and flexible way to the latter.

With the utilisation of external sources, our novel method allows turning points in the extrapolated hazard, which may provide a more flexible and realistic shape beyond the trial period. By adjusting relevant parameters of the weight function, the blended procedure permits non-monotonic hazards (as shown in Figure~\ref{haza}) that might be more practical in the extrapolation. For example, if a trial period ends with a low but increasing hazard, there could be several turning points over time, such as a temporary decrease due to the long-term survivors, then a following increase due to ageing effects \cite{latimer2022extrapolation}. Although the flexible parametric models, such as splines or fractional polynomials, can also capture a complex hazard function, a turning point cannot be generated in the post-trial period and the monotonic hazard based on the final observed segment is likely to be undesirable without external data.

It is important to identify appropriate external information to facilitate the extrapolation. A key assumption in the blended method is that from a specific time point the extrapolated survival is consistent with the estimate from external evidence. Before using potential data from another sources, researchers should examine if the external population matches some characteristics for the patients of interest and have equivalent mortality in the long-term. Conveniently, there is one advantage to the blending process that no adjustment would be required, even if 1:1 matching between two sources were unavailable. The matching procedure is replaced by the blending process. 

However, it should be noted that external patient-level data with even similar mechanism are not common in most situations. Therefore, we focus on more general cases such that only expert/clinical subjective beliefs are available in the long-term. Experts may have some knowledge about the likely values or plausible ranges of survival in the future according to the trial data and their experience. Previous applications of clinical opinion mainly involved validating the plausibility of extrapolations \cite{gibson2017modelling, bell2019review}, although a few studies combined the information into a model via a Bayesian framework, for example, with informative priors \cite{soikkeli2019extrapolating} or multiple parameter evidence synthesis \cite{guyot2017extrapolation}. Our approach could translate the beliefs about long-term survival to a representative curve, by interpreting the elicitation as an artificial data set, then fitting standard models to it. Meaningfully, the number of elicited time points is not limited and completely depends on the clinicians. Obviously, the curve would be closer to what the expert believes if more elicited information are collected. The procedure is simple and straightforward, yet the expert-based survival estimates are inherently subjective and maybe limited in scope, which means attention should be given to the selection of more appropriate knowledge if possible.

In the absence of long-term data within a trial itself, scenario-based sensitivity analyses should be performed for uncertain assumptions of the extrapolations. Uncertainty of the underlying evidence may have a large impact on the prediction. It may be worthwhile to test a range of plausible scenarios about the future trend, especially when integrating limited or conflicting elicitation into the extrapolations \cite{latimer2011nice}. This implementation is not hugely complicated, where the modeller simply changes values of the parameters associated with the blended model for defensible circumstances. In the extrapolated period, they can select suitable values (e.g., plausible ranges) of survival at multiple time points and flexibly determine the number of the elicited points and  locations. Besides, the blending operation, including the interval and the rate, can be characterised by the weight function if there is any available knowledge about biologically plausible shapes for the extrapolated hazard. A web-based application is being developed for aiding the elicitation process, in which immediate outcomes (i.e., survival and hazard plots) would help the experts to obtain reliable estimate. In future, there is possibility to expand the model to integrate  different kinds of elicited information, such as external hazard rates. 

Crucially, we believe that the blending method allows to shift elements of subjective assumptions away from the extrapolation derived from the observed data --- it is our view that untestable assumptions are all but unavoidable in the range of survival models that are relevant in HTA. The blending procedure attempts at recognising and embracing this feature, by providing a simple and powerful framework for its incorporation and evaluation on the model fit as well as on the long-term economic outputs.

There is no restrictive technical implementation for modelling the observed data. The piecewise Cox model is recommended due to the potential advantage of extremely good fits to the data without substantial computation required. Under a Bayesian framework, it allows a high level of flexibility and does not bring extra complexity, compared with spline or fractional polynomial models. Furthermore, the PH assumption is not necessary, as a stratified version of the Cox model exists, in which we can control for covariates that do violate the assumption by stratifying, effectively creating many versions of the baseline hazard. \cite{kleinbaum2012stratified}.

Current implications focus on the absolute effectiveness of treatments in a trial; however, decision making requires combination of different trials that compare multiple treatments with relative effectiveness of interest. In fact, it is not difficult to implement the blended approach into network meta-analysis (NMA) with a hierarchical structure that synthesises all direct and indirect evidence across trials \cite{caldwell2005simultaneous}. The mechanism of separately estimating observed and external hazards achieves flexibility in a simple way, and obtaining the blending result is not complex given a weight function identified by the experts. It is possible to apply common weight function with consistent values for the relevant parameters to all treatment arms, or alternatively consider different choices for each specific treatment if justified information is available externally. For the trial data, the Cox model would be beneficial, providing the same structure as other studies based on the PH assumption in the network. Moreover, interpretation of the parameters in Cox model is explicit. As the piecewise exponential model does not add much computation under Bayesian framework, the implementation of blending approach is less computationally intensive and therefore time consumption is probably less than alternative flexible models. 

\hypertarget{conclusion}{%
\section{Conclusion}\label{con}}

Long-term extrapolation entirely driven by the immature trial data is highly unreliable and varying assumptions of the treatment effect can have a great impact on the survival estimate. To improve the credibility of the prediction, the blended survival curve method allows the extrapolation taking advantage of external knowledge that manufacturers might have in form of hard data or just elicited belief from clinical experts. The formal inclusion of external evidence considers a variety of available sources, especially the subjective opinion that is more common in reality. Therefore, not only internal but also external validity can be fully taken into account for the survival model. Considering a range of plausible scenarios easily, the blended approach provides a simple and robust framework to ensure sufficient flexibility for the long-term survival estimate.

\end{document}